\documentclass[12pt]{article}

\usepackage{comment}
\usepackage{tikz}
\usepackage{caption}

\usepackage{mathtools}

\usepackage[mathscr]{euscript}

\usepackage{slashed}

\usepackage{amsmath}

\date{}

\title{\textbf{BFFT Nonlinear Constraints Abelianization of a
 Prototypical Second-Class System
}}
\author{ \textbf{
V. K. Pandey$^{a}$ and
R. Thibes$^{b}$}
\\\\
\textit{$^{a}$\small{Department of Physics and Astrophysics}}\\
\textit{\small{University of Delhi}}\\
\textit{\small{New Delhi -- 110007, India}}\\
\textit{$^{b}$\small{Departamento de Ci\^encias Exatas e Naturais}}\\
\textit{\small{Universidade Estadual do Sudoeste da Bahia}}\\
\textit{\small{Rodovia BR 415, Km 03, S/N, Itapetinga -- 45700-000, Brazil}}
 }

\usepackage{graphicx}
\usepackage{amssymb,amsfonts,amssymb,amsthm}
\usepackage{amsmath}

\begin{document}

\maketitle

\abstract{We apply the BFFT formalism to a prototypical second-class system, aiming to convert its constraints from second- to first-class.   The proposed system admits a consistent initial set of second-class constraints and an open potential function providing room for applications to mechanical models as well as field theory such as the non-linear sigma model.  The constraints can be arbitralily non-linear, broadly generalizing previously known cases.  We obtain a sufficient condition for which a simple closed expression for the Abelian converted constraints and modified involutive Hamiltonian can be achieved.  As explicit examples, we discuss a particle on a torus and a spontaneous Lorentz symmetry breaking vector model, obtaining in both situations the full first-class abelianized constraints in closed form and the corresponding involutive Hamiltonian.}

\section{Introduction}
Since the foundational works of Anderson, Bergmann and Dirac \cite{Dirac:1950pj, Anderson:1951ta, Dirac}, the quantization of constrained dynamical systems has been extensively studied in the theoretical physics literature from many different perspectives.  A successful path to quantizing systems with second-class constraints through their conversion to first-class, using auxiliary variables, has been pointed out along the groundbreaking references \cite{Batalin:1986aq, Batalin:1986fm, Egorian:1988ss, Batalin:1989dm, Batalin:1991jm}, culminating in what came to be known as the Batalin-Fradkin-Fradkina-Tyutin (BFFT)\footnote{Also known in the literature as BFT, BF or BT.} constraints abelianization formalism.  The main idea of the BFFT scheme comes to converting all second-class contraints to first-class, in order to benefit from the existing powerfull techniques and tools for the quantization of gauge-invariant systems, such as the BFV \cite{Fradkin:1975cq, Batalin:1977pb, Batalin:1983pz}, as well as its Lagrangian counterpart field-antifield BV \cite{Batalin:1981jr, Batalin:1984jr, Batalin:1985qj} methods.  In the present paper, we shall be concerned with the application of the BFFT formalism to a prototypical second-class system with nonlinear constraints, to be defined below in Section {\bf 2}, which encompasses a huge class of important physical models, promoting a broader general view and understanding of the mentioned formalism and providing significant technical shortcuts for its practical applications.

The current physical understanding of nature's fundamental interactions certainly passes through gauge symmetry.  In fact, all major models in quantum field theory enjoy gauge-invariance properties.  Upon quantization, that symmetry leads to the important Becchi-Rouet-Stora-Tyutin (BRST) transformations \cite{Becchi:1974xu, Becchi:1974md, Tyutin:1975qk, Becchi:1975nq} mixing physical and ghost fields, producing the conserved BRST charge and allowing the investigation of deep consistency internal results and interelations such as its renormalizability and physical content.  As dynamical Hamiltonian systems \cite{Sundermeyer:1982gv, Gitman:1990qh, Henneaux:1992ig}, gauge theories are necessarily constrained in the sense of Dirac \cite{Dirac:1950pj, Anderson:1951ta, Dirac}, possessing non-physical degrees of freedom.   The mentioned constraints are mathematical relations satisfied by the phase space variables and can be classified into first- and second- classes according to their Poisson bracket structure.\footnote{By definition, a first-class phase space function has a vanishing Poisson bracket within the constraints hypersurface with all constraints.  Otherwise, the function is classified as second-class.}  A key fact resulting from the Dirac-Bergmann analysis is that only first-class constraints can generate gauge symmetries.  As it is well-known, in principle the canonical quantization of systems with second-class constraints can be achieved by replacing the usual Poisson brackets by the Dirac ones.  In that case, in the absence of gauge-invariance, no gauge-fixing is needed. However, for field theory systems of physical interest, the Dirac bracket structure produces cumbersome nonlocal relations among the quantized fields turning that canonical approach into a technical difficult task.  In this way, the idea of converting second-class constraints to first-class flourished as a much more natural and achievable alternative.  Prior to the appearence and consolidation of the BFFT systematical approach, early accounts on that idea can be seen in references \cite{Stueckelberg:1957zz, Wess:1971yu, Faddeev:1986pc} where auxiliary variables were introduced by hand to convert specific constraints to first-class.
Aiming at generality, Batalin and Fradkin first published an operatorial version for an overall canonical quantization method transforming the second-class constraints into first-class introducing convenient auxiliary fields \cite{Batalin:1986aq, Batalin:1986fm}.  A complementary approach introducing classical auxiliary variables at an earlier step in phase space was then proposed by Egorian and Manvelyin in reference \cite{Egorian:1988ss} and later shown by Batalin, Fradkin and Fradkina \cite{Batalin:1989dm} to be equivalent to the original treatment of \cite{Batalin:1986aq, Batalin:1986fm}.  Finally, Batalin and Tyutin settled the case proving an existence theorem for the solution of the general problem of Abelian conversion of second-class constraints \cite{Batalin:1991jm}.  A recent review with a compact reformulation of the BFFT conversional approach can be seen in \cite{Batalin:2018wxh}.
 
Seminal applications of the BFFT formalism can be seen in references  \cite{Fujiwara:1989ia, Kim:1992ey, Banerjee:1993ae, Banerjee:1993ac, Amorim:1994np, Kim:1994rh, Amorim:1994ft, Kim:1996gk}, while more recent accounts can be found in \cite{Taie:2014oza, Sararu:2016qnx, Gharavi:2016nlv, Ghosal:2020syu}.  In \cite{Taie:2014oza}, the authors show an interesting study of graphene in non-commutative space in which the BFFT formalism is used to recover broken gauge symmetries, whereas \cite{Sararu:2016qnx} deals with a constrained higher-order-derivatives model within the BFFT conversional approach.  A direct application of the BFFT formalism to the non-Abelian supersymmetric Chern-Simons model can be seen in reference \cite{Gharavi:2016nlv}.  In reference \cite{Ghosal:2020syu}, we can see the use of BFFT variables in a constraints conversional approach to discuss the quantization of the anomalous chiral Schwinger model. It is also worth mentioning here the appearance of many extensions and generalizations of the original BFFT formalism such as the improved BFFT \cite{Kim:1997psa, Park:1997zj}, the embedding BFFT \cite{Banerjee:1994pp, Oliveira:1998ek, Abreu:2000ip}, the Wotsazek-Neves \cite{Neves:1999jr, Abreu:2000sq} and the gauge-unfixing \cite{Anishetty:1992yk, Vytheeswaran:1994np, Neto:2006gt, Neto:2009rm, Monemzadeh:2014wma} methods.
Since the BFFT general solution, resulting from the analysis of references \cite{Batalin:1986aq, Batalin:1986fm, Egorian:1988ss, Batalin:1989dm, Batalin:1991jm}, comes as a power series in the auxiliary variables, the question of obtaining closed expressions for the converted constraints poses itself as a natural one.  For the particular case when the original constraint equations are linear in the original variables, it has been shown in \cite{Amorim:1994ua, Amorim:1995sh, Amorim:1995mr} that the Abelianization process corresponds to a linear shift in the original phase space variables.  On the other hand, the case of non-linear constraints has been discussed for instance in \cite{BarcelosNeto:1997qk, Dehghani:2014doa}, but still lacks a final conclusive answer with respect to a general and practical closed form. 
 
Looking for patterns and searching for uniformity in the BFFT and its modified approaches to a wide range of second-class models in the literature, we propose a prototypical second-class dynamical system which includes many of those models as particular cases.  Our prototype system generalizes a previous Hamiltonian model recently published in \cite{Thibes:2020yux} by allowing a larger set of constraints and including a generic open potential function.
The analysis carried out in reference \cite{Thibes:2020yux} was not directly related to the BFFT formalism.  Although with a seemingly different approach, the simplified Hamiltonian model of \cite{Thibes:2020yux} was shown to possess gauge-invariance, being suitable for a consistent functional quantization with BRST symmetry.  In the present paper, we considerably extend that model in order to properly use the BFFT formalism.
A set of initial constraint functions is included in the starting Lagrangian by means of Lagrange multipliers, which are nevertheless considered as ordinary variables.  The application of the Dirac-Bergmann algorithm produces four families of independent constraints in phase space and, under a consistency assumption, the whole set of constraints is shown to be second-class.  We proceed with the BFFT formalism introducing one auxiliary variable for each second-class constraint, pursuing the full abelianization of the constraints algebra.

Our work is organized as follows.  In Section {\bf 2} below, we introduce the main prototypical system in Lagrangian form, defining our notation and conventions, and show that it characterizes a second-class system.  In Section {\bf 3}, we briefly review the central ideas of the BFFT formalism contextualized along our prototypical system and derive a suficient condition to obtain a closed expression for the converted constraints and Hamiltonian.  In Section {\bf 4}, we apply the developed technique to a mechanical model, namely, the Abelianization of the constraints describing the motion of a particle on a torus surface.  In Section {\bf 5}, we explicitly show that the proposed prototypical system can be readily applied to field theory and discuss the Abelianization of a Lorentz symmetry breaking bumblebee model.  We close in Section {\bf 6} with our conclusion and final remarks.
 
\section{A Prototypical Second-Class System}
Given a symmetric invertible square matrix $f_{ij}(q^k)$ depending on the variables $q^k$, with $i, j, k = 1, \dots, N$, and $M$ thrice differentiable functions $T_\alpha(q^k)$, with $\alpha=1,\dots,M$, we define our prototypical system by the Lagrangian\footnote{Along the text, an upper and lower repeated index always means summation throughout the corresponding range.  For instance we have $f_{ij}(q^k){{\dot{q}}^i}{{\dot{q}}^j}\equiv\displaystyle\sum_{i,j=1}^{N}f_{ij}(q^k){{\dot{q}}^i}{{\dot{q}}^j}$ as well as $l^\alpha T_\alpha(q^k)\equiv\displaystyle\sum_{\alpha=1}^{M}l^\alpha T_\alpha(q^k)$, so on and so forth.}
\begin{equation}\label{L}
L(l^\alpha,q^k,{\dot{q}}^k)=\frac{1}{2}f_{ij}(q^k){{\dot{q}}^i}{{\dot{q}}^j}
-V(q^k)
-l^\alpha T_\alpha(q^k)
\,.
\end{equation}
Although the variables $l^\alpha$ in a certain sense enter in (\ref{L}) as Lagrange multipliers, at this point they are considered as configuration space variables in very much the same level as $q^k$.
As usual, a dot placed over a variable indicates its derivative with respect to the time evolution parameter.

The Hamiltonian version of (\ref{L}) corresponds to a constrained dynamical system in phase space.  In fact, introducing momenta variables canonically conjugated to $q^i$ and $l^\alpha$ defined respectively as
\begin{equation}
p_i\equiv \frac{\partial L}{\partial {\dot{q}}^i}\,\, 
\mbox{ and }\,\,
\pi_\alpha \equiv \frac{\partial L}{\partial {\dot{l}}^\alpha}
\,,
\end{equation}
we obtain the canonical Hamiltonian
\begin{equation}\label{H}
H=\frac{1}{2}f^{ij}(q^k)p_ip_j+V(q^k)+l^\alpha T_\alpha(q^k)
\end{equation}
and a first set of $M$ trivial primary constraints
\begin{equation}\label{Xi1}
\chi_{(1)\alpha} = \pi_\alpha\,,\,\,\,\,\alpha=1,\dots,M
\,.
\end{equation}
The upper index functions $f^{ij}(q^k)$ introduced in the Hamiltonian
(\ref{H}) denote the inverse of the previous lower index ones $f_{ij}(q^k)$, in other words we have
\begin{equation}
f^{ik}f_{kj}=\delta^i_j
\end{equation}
by definition.
For the next algebraic manipulations, we shall often need the partial derivatives of the functions $f^{ij}(q^k)$, $V(q^k)$ and $T_\alpha(q^k)$, for that matter we introduce the condensed brief notations
\begin{equation}
f^{ij}_{\,\,\,\,,k}\equiv \frac{\partial f^{ij}}{\partial q^k}
\,,~~~~~
f^{ij}_{\,\,\,\,,kl}\equiv \frac{\partial^2 f^{ij}}{\partial q^l\partial q^k}
\,,~~~~~
\mbox{etc,}
\end{equation}
\begin{equation}\label{Vi}
V_{i}\equiv\frac{\partial V}{\partial q^i}\,,~~~~~V_{ ij}\equiv\frac{\partial^2 V}{\partial q^j \partial q^i}\,,~~~~~\mbox{etc,}
\end{equation}
and
\begin{equation}\label{Ti}
T_{\alpha i}\equiv\frac{\partial T_\alpha}{\partial q^i}\,,~~~~~T_{\alpha ij}\equiv\frac{\partial^2 T_\alpha}{\partial q^j \partial q^i}\,,~~~~~\mbox{etc.}
\end{equation}

Following further the Dirac-Bergmann algorithm \cite{Dirac:1950pj, Anderson:1951ta, Dirac}, we impose the stability of the constraints under time evolution.  In this way, additionally to (\ref{Xi1}), three more constraint families are generated, namely,
\begin{equation}\label{Xi2}
\chi_{(2)\alpha}=T_\alpha
\,,
\end{equation}
\begin{equation}\label{Xi3}
\chi_{(3)\alpha} = f^{ij}p_iT_{\alpha j}
\,,
\end{equation}
and
\begin{equation}\label{Xi4}
\chi_{(4)\alpha} = \frac{1}{2}Q_\alpha^{ij}p_ip_j
-v_\alpha-l^\beta w_{\alpha\beta}
\,,
\end{equation}
constituting a total of $4M$ constraints in phase space.
For notation convenience, in the RHS of equation (\ref{Xi4}) we have introduced the $q^k$-dependent
quantities $Q^{ij}_\alpha$, $v_\alpha$ and $w_{\alpha \beta}$  defined explicitly by
\begin{equation}\label{Qij}
Q^{ij}_\alpha\equiv
{\left(f^{il}T_{\alpha l}\right)}_{,k} f^{kj}
+{\left(f^{jl}T_{\alpha l}\right)}_{,k} f^{ki}
-f^{kl}T_{\alpha k} f^{ij}_{\,\,\,\,,l}
\,,
\end{equation}
\begin{equation}\label{v}
v_\alpha\equiv f^{ij}T_{\alpha i}V_j
\,
\end{equation}
and
\begin{equation}\label{w}
w_{\alpha\beta}\equiv f^{ij}T_{\alpha i} T_{\beta j}
\,.
\end{equation}

As it turns out, if we assume
\begin{equation}\label{assumption}
w\equiv \det w_{\alpha\beta} \neq 0
\,,
\end{equation}
the whole set of constraints $\chi_{(r)\alpha}$ with $r=1,\dots,4$ is second-class.  This can be seen by
computing the Poisson brackets among all constraints and writing the resulting constraint matrix as 
\begin{equation}\label{CM}
\Delta_{(rs)\alpha\beta}\equiv\lbrace\chi_{(r)\alpha}, \chi_{(s)\beta} \rbrace =
\displaystyle
\left[
\begin{array}{cccc}
0 & 0 & 0 & w_{\alpha\beta} \\ 
0 & 0 &w_{\alpha\beta} &D_{\alpha\beta} \\
0 & -w_{\alpha\beta}  & M_{\alpha\beta} & R_{\alpha\beta} \\
-w_{\alpha\beta}   & -D_{\beta\alpha} & -R_{\beta\alpha} & N_{\alpha\beta}
\end{array}
\right]
\,,
\end{equation}
for $r,s=1,\dots,4$, with the short-hand conventions
\begin{equation}
D_{\alpha\beta}\equiv
\lbrace
\chi_{(2)\alpha}, \chi_{(4)\beta}
\rbrace
\,,\,\,\,\,
M_{\alpha\beta}\equiv
\lbrace
\chi_{(3)\alpha}, \chi_{(3)\beta}
\rbrace
\,,\,\,\,\,
N_{\alpha\beta}\equiv
\lbrace
\chi_{(4)\alpha}, \chi_{(4)\beta}
\rbrace
\,,
\end{equation}
and
\begin{equation}
R_{\alpha\beta}\equiv
\lbrace
\chi_{(3)\alpha}, \chi_{(4)\beta}
\rbrace
\,.
\end{equation}
In fact, the determinant of the constraint matrix (\ref{CM}) depends only on its secondary diagonal and is given by
\begin{equation}
\det \Delta_{(rs)\alpha\beta} =
w^4
\end{equation}
which clearly shows that, under the assumption (\ref{assumption}), the prototypical system (\ref{L}) is indeed second-class.

For completeness and future reference, we compute the remaining further entries below the main diagonal in the constraint matrix (\ref{CM}) and write them explicitly as 
\begin{equation}\label{D}
D_{\alpha\beta}=T_{\alpha i}Q^{ij}_\beta p_j
\,,
\end{equation}
\begin{equation}
M_{\alpha\beta}=f^{ij}p_k
\left[
{\left(f^{kl}T_{\alpha l}\right)}_{,i}T_{\beta j}
-
{\left(f^{kl}T_{\beta l}\right)}_{,i}T_{\alpha j}
\right]
=
p_k M^k_{\alpha\beta}
\,,
\end{equation}
\begin{equation}
N_{\alpha\beta}= p_i
\left[
\chi_{(4)\alpha,j}Q^{ij}_\beta
-
\chi_{(4)\beta,j}Q^{ij}_\alpha
\right]
=
p_i p_j p_k Q^{ijk}_{\alpha\beta}
+p_i V^i_{\alpha\beta}+p_il^\gamma N^i_{\alpha\beta\gamma}
\,,
\end{equation}
and
\begin{equation}\label{R}
R_{\alpha\beta}
=
p_i p_j R^{ij}_{\alpha\beta}+ f^{ij}T_{\alpha i} v_{\beta j} 
+ l^\gamma f^{ij}T_{\alpha i} w_{\beta\gamma j}
\,,
\end{equation}
with the definitions
\begin{equation}\label{D1}
M^k_{\alpha\beta}\equiv f^{ij}
\left[
\left(f^{kl}T_{\alpha l}\right)_{,i}T_{\beta j}
-
\left(f^{kl}T_{\beta l}\right)_{,i}T_{\alpha j}
\right]
\,,
\end{equation}
\begin{equation}
Q^{ijk}_{\alpha\beta}\equiv\frac{1}{2}
\left(
Q^{ij}_{\alpha,l}Q^{kl}_\beta
-
Q^{ij}_{\beta,l}Q^{kl}_\alpha
\right)
\,,
\end{equation}
\begin{equation}
V^i_{\alpha\beta}\equiv
v_{\alpha j} Q^{ij}_\beta
-
v_{\beta j} Q^{ij}_\alpha
\,,
\end{equation}
\begin{equation}
N^i_{\alpha\beta\gamma}\equiv
w_{\alpha\gamma j} Q^{ij}_\beta
-
w_{\beta\gamma j} Q^{ij}_\alpha
\end{equation}
and
\begin{equation}\label{D5}
R^{ij}_{\alpha\beta}\equiv
{\left(f^{ik}T_{\alpha k}\right)}_{,l}Q^{lj}_\beta
-\frac{1}{2}f^{kl}T_{\alpha k} Q^{ij}_{\beta l}
\,.
\end{equation}
The functions  $v_{\alpha i}$ and $w_{\alpha\beta i}$ denote the corresponding derivatives of (\ref{v}) and (\ref{w}) with respect to $q^i$, in the same vein as equations (\ref{Vi}) and (\ref{Ti}).  
Note that all the above defined quantities, along equations (\ref{D1}) to (\ref{D5}), depend only on the coordinates $q^k$.

As mentioned in the Introduction, in spite of its simplicity, the prototypical system (\ref{L}) can describe many models of physical interest.  In particular, the index sums can be straightforwardly generalized to integrations in field theory models.  In reference \cite{Thibes:2020yux}, a Hamiltonian related to a simplified version of (\ref{L}) has been used to study the BRST symmetries of the nonlinear $O(N)$ sigma model.  In the next section, we briefly review the main idea of the BFFT abelianization procedure and apply it to the prototypical system (\ref{L}), aiming to convert the $4M$ constraints (\ref{Xi1}) and (\ref{Xi2}) to (\ref{Xi4}) to first-class.

\section{BFFT Abelianization Procedure}
The Batalin-Fradkin-Fradkina-Tyutin quantization approach
\cite{Batalin:1986aq, Batalin:1986fm, Egorian:1988ss, Batalin:1989dm, Batalin:1991jm}
aims to convert the second-class constraints, in our case all $\chi_{(r)\alpha}$, to corresponding first-class ones by extending the initial phase space $(q^k,l^\alpha,p_k,\pi_\alpha)$ including new BFFT variables $\eta^{(r)\alpha}$.  In order to preserve the number of degrees of freedom along the conversion process, we introduce precisely one BFFT variable for each original second-class constraint.  Hence, for the present prototypical system, the indexes run through $r=1,\dots,4$, $\alpha=1,\dots,M$, and we have a total of $4M$  $\eta^{(r)\alpha}$ BFFT variables.  The sought converted constraints ${\tilde\chi}_{(r)\alpha}$ can be expanded in a power series in the auxiliary BFFT variables as
\begin{equation}\label{Xitilde}
{\tilde\chi}_{(r)\alpha} = \sum_{n=0}^{\infty}X_{(rt_{(n)})\alpha \gamma_{(n)}}\eta^{(t_{(n)})\gamma_{(n)}}
\,.
\end{equation}
We are using here the compact notation introduced in reference \cite{Thibes:2020bkk}, by which equation (\ref{Xitilde}) above more explicitly means
\begin{align}\label{Xtilde2}
{\tilde\chi}_{(r)\alpha} =& X_{(r)\alpha}
+X_{(rt_{1})\alpha \gamma_{1}}\eta^{(t_{1})\gamma_{1}}
+X_{(rt_{1}t_{2})\alpha \gamma_{1}\gamma_{2}}\eta^{(t_{1})\gamma_{1}} \eta^{(t_{2})\gamma_{2}} \nonumber\\&
+X_{(rt_{1}t_{2}t_{3})\alpha \gamma_{1}\gamma_{2}\gamma_{3}}\eta^{(t_{1})\gamma_{1}} \eta^{(t_{2})\gamma_{2}}  \eta^{(t_{3})\gamma_{3}} + \dots
\,,
\end{align}
with usual further summations for repeated indexes.\footnote{For instance, the second term on the RHS of (\ref{Xtilde2}) stands for
$$
X_{(rt_{1})\alpha \gamma_{1}}\eta^{(t_{1})\gamma_{1}}=
\sum_{\gamma_1=1}^{M}\left[
X_{(r1)\alpha\gamma_1}\eta^{(1)\gamma_1}+
X_{(r2)\alpha\gamma_1}\eta^{(2)\gamma_1}+
X_{(r3)\alpha\gamma_1}\eta^{(3)\gamma_1}+
X_{(r4)\alpha\gamma_1}\eta^{(4)\gamma_1}
\right]\,\,\mbox{ etc.}
$$}
The coefficients $X_{(rt_{(n)})\alpha \gamma_{(n)}}$, with $n\in\mathbb{N}$, are functions of the original phase space variables, i.e.,
\begin{equation}
X_{(rt_{(n)})\alpha \gamma_{(n)}} = X_{(rt_{(n)})\alpha \gamma_{(n)}}
(q^k,l^\alpha,p_k,\pi_\alpha)
\,,
\end{equation}
to be determined in such a way that (\ref{Xitilde}) are indeed first-class.

The converted first-class constraints ${\tilde\chi}_{(r)\alpha}$ should coincide with the original second-class ones when all the BFFT variables are set to zero.  This immediately leads, for $n=0$, to
\begin{equation}\label{bc}
X_{(r)\alpha} = \chi_{(r)\alpha}
\,.
\end{equation}
Concerning the higher-order coefficient functions present in (\ref{Xitilde}), we impose the Abelianization relations
\begin{equation}\label{Abel}
\{ {\tilde\chi}_{(r)\alpha} , {\tilde\chi}_{(s)\beta}  \} = 0
\end{equation}
and the analysis of the resulting expression order by order in the BFFT variables $\eta^{(r)\alpha}$ leads to a set of conditions for $X_{(rt_{(n)})\alpha \gamma_{(n)}}$, $n>0$.
Actually, the requirement (\ref{Abel}) is stronger than one needs to obtain first-class constraints, as it implies the identical vanishing of the Poisson brackets among the converted constraints with a consequent Abelian constraints algebra.   Substituting (\ref{Xitilde}) into (\ref{Abel}) and picking up the resulting $\eta^{(r)\alpha}$ zero-order term, it is straightforward to obtain the well-known BFFT first-step-condition 
\begin{equation}\label{BFFT1}
 \Delta_{(rs)\alpha\beta}= 
 - X_{(rt_{1})\alpha \gamma_{1}}
 \omega^{(t_{1}t_{2})\gamma_{1}\gamma_{2}}
 X_{(st_{2})\beta \gamma_{2}}
\,,
\end{equation}
where $\omega^{(t_{1}t_{2})\gamma_{1}\gamma_{2}}$, defined as
\begin{equation}
\omega^{(rs)\alpha\beta}\equiv \{ \eta^{(r)\alpha} , \eta^{(s)\beta} \}
\,,
\end{equation}
characterizes the symplectic structure among the BFFT variables.  In practical terms, one chooses a convenient $\omega^{(rs)\alpha\beta}$ and solves (\ref{BFFT1}) for $X_{(rs)\alpha \beta}$.  In the particular case of linear second-class constraints, it is sufficient to solve (\ref{BFFT1}) as the higher-order $\eta^{(r)\alpha}$ in (\ref{Xitilde}) are null.  Furthermore in that case, it has been shown in references \cite{Amorim:1994ua, Amorim:1995sh, Amorim:1995mr} that the BFFT abelianization can be extended to all phase space functions in terms of a shift in the original phase space variables by the BFFT ones.  On the other hand, nonlinear constraints, as can be the situation for our $\chi_{(r)\alpha}$, are more tricky and demand a more careful treatment \cite{BarcelosNeto:1997qk, Dehghani:2014doa} -- after attending (\ref{BFFT1}) it is necessary to move on to consider higher-order terms in the expansion (\ref{Xitilde}).   The existence of solutions for (\ref{Abel}) as a power series in the BFFT variables of the form (\ref{Xitilde}) for the general case has been established in reference \cite{Batalin:1991jm} by obtaining recursive relations for the functions $X_{(rt_{(n)})\alpha \gamma_{(n)}}\eta^{(t_{(n)})\gamma_{(n)}}$ for arbitrary $n$.

For our current prototypical second-class system, aiming to preserve the initial constraints original structure as much as possible,
we choose the symplectic BFFT algebra
\begin{equation}\label{omegars}
\omega^{(rs)\alpha\beta}=
\displaystyle
\left[
\begin{array}{cccc}
0 & 0 & \delta^{\alpha\beta} & 0 \\ 
0 & 0 & 0 & \delta^{\alpha\beta} \\
-\delta^{\alpha\beta} & 0  & 0 & 0 \\
0 & -\delta^{\alpha\beta} & 0 & 0
\end{array}
\right]
\end{equation}
and consider a deformation in phase space induced by the BFFT variables corresponding to the ansatz\footnote{For notation convenience, we redefine $\eta^{(2)}_\alpha\equiv\eta^{(2)\alpha}$ and $\eta^{(3)}_\alpha\equiv\eta^{(3)\alpha}$ so that the non-null Poisson brackets among the BFFT variables can be written as $\lbrace\eta^{(2)}_\alpha, \eta^{(4)\beta}\rbrace=\lbrace\eta^{(1)\alpha}, \eta^{(3)}_\beta\rbrace=\delta^\alpha_\beta$.}
\begin{equation}\label{chitilde}
\begin{aligned}
{\tilde\chi}_{(1)\alpha}&=\pi_\alpha - \eta^{(3)}_\alpha
\,,\\
{\tilde\chi}_{(2)\alpha}&=T_\alpha + \eta^{(2)}_\alpha
\,,\\
{\tilde\chi}_{(3)\alpha}&= \bar{f}^{ij}
\left(
p_i-T_{\beta i}\eta^{(4)\beta}
\right)
\bar{T}_{\alpha j}
\,,
\end{aligned}
\end{equation}
and
\begin{equation}\label{chitilde4}
{\tilde\chi}_{(4)\alpha}=
\frac{1}{2}
{\bar Q}_\alpha^{ij}\left(p_i-T_{\beta i}\eta^{(4)\beta}\right)\left(p_j-T_{\gamma j}\eta^{(4)\gamma}\right)
-{\bar v}_\alpha - l^\beta \bar{w}_{\alpha\beta}
-\bar{w}_{\alpha\beta}\eta^{(1)\beta}
\end{equation}
with all barred quantities depending only on the variables $q^k$ and $\eta^{(2)}_\alpha$.
In other words, in equations (\ref{chitilde}) we have
\begin{equation}\label{fbar}
\bar{f}^{ij}=\bar{f}^{ij}(q^k,\eta^{(2)}_\alpha)
\,\,\,\,\,
\mbox{ and }
\,\,\,\,\,
\bar{T}_{\alpha i}=\bar{T}_{\alpha i}(q^k,\eta^{(2)}_\alpha)
\end{equation}
as well as, in equation (\ref{chitilde4}),
\begin{equation}\label{Qbar}
{\bar Q}_\alpha^{ij}={\bar Q}_\alpha^{ij}(q^k,\eta^{(2)}_\alpha)
\,,\,\,\,\,\,
\bar{v}_{\alpha}=\bar{v}_{\alpha}(q^k,\eta^{(2)}_\alpha)
\,,\,\,\,\,\,
\mbox{ and }
\,\,\,\,\,
\bar{w}_{\alpha\beta}=\bar{w}_{\alpha\beta}(q^k,\eta^{(2)}_\alpha)
\,.
\end{equation}

Due to the boundary condition (\ref{bc}), the barred quantities above should reproduce the corresponding unbarred ones in the limit $\eta^{(2)\alpha}\rightarrow0$.  That means, if $\bar{F}(q^k,\eta^{(2)}_\alpha)$ represents any of the functions in equations (\ref{fbar}) or (\ref{Qbar}), we should have
\begin{equation}
\bar{F}(q^k,0)=F(q^k)
\,.
\end{equation}
In this way, using the compact notation of \cite{Thibes:2020bkk}, we may expand $\bar{F}$ in Taylor series in the BFFT variable $\eta^{(2)}_\alpha$ as
\begin{equation}\label{Fbar}
\bar{F}(q^k,\eta^{(2)}_\alpha) = \sum_{n=0}^{\infty}\, \frac{F^{\alpha_{(n)}}}{n!}\,\eta^{(2)}_{\alpha_{(n)}}
\end{equation}
with
\begin{equation}
F^{\alpha_{(n)}} = F^{\alpha_{(n)}}(q^k)
\,.
\end{equation}
Now it is straightforward to check that the converted constraints (\ref{chitilde}) and (\ref{chitilde4}) will generate the Abelian algebra
(\ref{Abel}) if the barred quantities (\ref{fbar}) and (\ref{Qbar}), generically represented by $\bar{F}(q^k,\eta^{(2)}_\alpha)$, satisfy
the condition
\begin{equation}\label{c1}
\bar{F}_{,i}=T_{\alpha i}\lbrace \bar{F}, \eta^{(4)\alpha} \rbrace
\,,
\end{equation}
with $,i$ denoting partial derivative with respect to $q^i$.
Inserting the expansion (\ref{Fbar}) into (\ref{c1}) leads to relation
\begin{equation}\label{r1}
F^{\alpha_{(n)}}_{,i}=T_{\alpha i} F^{\alpha\alpha_{(n)}}\,,
\end{equation}
constraining the components of the Taylor expansion of (\ref{Fbar}) among themselves and the derivatives of the functions $T_\alpha$.  If (\ref{r1}) is satisfied for all $i$ and $n$, we may integrate it out and solve for $F^{\alpha_{(n)}}$.  For that matter, introduce $w^{\alpha\beta}$, the inverse matrix of (\ref{w}) satisfying
\begin{equation}
w^{\alpha\gamma}w_{\gamma\beta}=\delta^\alpha_\beta
\,,
\end{equation}
and note that the multiplication of (\ref{r1}) by $f^{ij}T_{\alpha j}$, with a corresponding sum in the index $i$, produces the well-defined recursive solution
\begin{equation}\label{recsol}
F^{\alpha\alpha_{(n)}}=f^{ij}w^{\alpha\beta}T_{\beta i}F^{\alpha_{(n)}}_j
\,.
\end{equation}
We remark that the inverse $w^{\alpha\beta}$ exists precisely due to the assumption (\ref{assumption}).

In order to complete the Abelianization process, we point out that the corresponding modified Hamiltonian in strong involution with the converted constraints can be directly obtained from (\ref{H}), considering the same deformation in phase space introduced in equations (\ref{chitilde}) and (\ref{chitilde4}), as 
\begin{eqnarray}\label{Htilde}
\tilde{H}&=&\frac{1}{2}\bar{f}^{ij}(q^k,\eta^{(2)}_{\alpha})\left(p_i-T_{\beta i}\eta^{(4)\beta}\right)\left(p_j-T_{\gamma j}\eta^{(4)\gamma}\right)
\nonumber\\&&
+\,\bar{V}(q^k,\eta^{(2)}_{\alpha})+(l^\alpha+\eta^{(1)\alpha})\left( T_\alpha(q^k) + \eta^{(2)}_\alpha \right)
\,,
\end{eqnarray}
with
\begin{equation}\label{c2}
\bar{V}_{,i}=T_{\alpha i}\lbrace \bar{V}, \eta^{(4)\alpha} \rbrace
\,.
\end{equation}

Therefore we see that (\ref{c1}) provides a sufficient condition for a simple solution of the constraints Abelianization problem (\ref{Abel}), whilst the additional relation (\ref{c2}) allows for a corresponding simple involutive Hamiltonian.  Namely, under (\ref{c1}), the converted constraints (\ref{chitilde}) and (\ref{chitilde4}) represent simple closed expressions in the BFFT variables $\eta^{(1)\alpha}, \eta^{(3)}_\alpha, \eta^{(4)\alpha}$ containing a Taylor series in $\eta^{(2)}_\alpha$ of the form (\ref{Fbar}) with coefficients (\ref{recsol}).  A corresponding statement holds for the existence of a simple involutive Hamiltonian in the form (\ref{Htilde}) satisfying
\begin{equation}
\{  {\tilde\chi}_{(r)\alpha}, \tilde{H} \} = 0\,,\,\,\,r=1,\dots,4,\,\,\,\alpha=1,\dots,M
\,,
\end{equation}
provided the potential function admits an extension $\bar V$ satisfying (\ref{c1}). In the next sections, we apply the general ideas discussed above to two particular specific models.

\section{Mechanical Example:  Particle on a torus}
In this section we illustrate the application of the prototypical Lagrangian (\ref{L}) to describe a particle constrained to move on a torus surface as a second-class system and perform its Abelianization.  Our goal is to obtain and exhibit the converted Abelian first-class constraints and modified involutive Hamiltonian in closed form for this particular example to clarify the ideas introduced in the last two sections.
The particle on a torus has been recently discussed in reference \cite{Pandey:2017kwc} through the finite field dependent BRST formalism \cite{Joglekar:1994tq, Banerjee:2000jt, Upadhyay:2012qw}, exploring the BRST symmetry generators algebra.  Other aspects of this model, concerning its canonical quantization, energy spectrum, superfield approach and BRST symmetries, can be seen in \cite{Hong:2005qom, Xun:2013qom, Kumar:2014fca}.
The Lagrangian for a mass $m$ particle confined to move along a two-dimensional toric surface embedded in three dimensions can be written as
\begin{eqnarray}
L = \frac{1}{2} m {\dot r}^2+\frac{1}{2} m r^2 {\dot \theta}^2+\frac{1}{2}m (b+r \sin\theta)^2{\dot\phi}^2 - l (r-a)
\,,
\label{lagft}
\end{eqnarray}
with $a$ and $b$ denoting respectively the torus minor and major radii and $(r,\theta,\phi)$ toroidal coordinates related to the usual Cartesian ones $(x,y,z)$ as
\begin{equation}
x = (b+r\sin\theta)\cos\phi,
\quad y = (b+r\sin\theta)\sin\phi,
\quad z = r\cos\theta
\,.
\end{equation}
In principle $(r,\theta,\phi)$ can vary freely in three dimensions, but then comes the extra variable $l$ in (\ref{lagft}) whose equation of motion constrains $r$ to equal $a$.  Confronting the current Lagrangian function (\ref{lagft}) to the general prototypical one in equation (\ref{L}), we have $N=3$, $M=1$, an $(r,\theta)$-dependent metric function with diag$(f_{ij})=m(1,r^2,(b+r\sin\theta)^2)$, a null potential function and
\begin{equation}\label{T=r-a}
T(r,\theta,\phi)=r-a\,.
\end{equation}

Introducing the canonical momenta $p_r, p_\theta, p_\phi$ and $\pi$,  respectively conjugated to the coordinates $r, \theta, \phi$ and $l$, the canonical Hamiltonian corresponding to the Lagrangian (\ref{lagft}) can be written as
\begin{eqnarray}
H = \frac{p^2_r}{2m}+\frac{p^2_\theta}{2mr^2}+\frac{p^2_\phi}{2m(b+r\sin\theta)^2} + l (r-a)
\label{Ham}
\,,
\end{eqnarray}
with the inverse of $f_{ij}$ given by
\begin{equation}\label{fij}
f^{ij}\equiv\left[
\begin{array}{ccc}
f^{rr}&0&0\\
0&f^{\theta\theta}&0\\
0&0&f^{\phi\phi}
\end{array}
\right]
=
\frac{1}{m}
\left[
\begin{array}{ccc}
1&0&0\\
0&r^{-2}&0\\
0&0&{(b + r \sin\theta)^{-2}}
\end{array}
\right]
\,.
\end{equation}
Since $\dot l$ does not show up in (\ref{lagft}), we immediately have $\chi_{(1)}=\pi$ as a primary constraint.
After computing $Q^{ij}$ from equation (\ref{Qij}) as
\begin{equation}\label{Qijtorus}
Q^{ij} = 
\left[
\begin{array}{ccc}
0 & 0 & 0\\
0 & \displaystyle\frac{2}{m^2r^3} & 0\\
0 & 0 & \displaystyle\frac{2\sin\theta}{m^2(b+r\sin\theta)^3}
\end{array}
\right]
\,,
\end{equation}
the remaining second-class constraints can be directly read from (\ref{Xi2}) to (\ref{Xi4}) as\footnote{Note from (\ref{T=r-a}) that $T_r=1$ and $T_\theta=T_\phi=0$.}
\begin{equation}
\begin{aligned}
\chi_{(2)} &= T = r-a\,, \\
\chi_{(3)} &= f^{rr}p_r T_r + f^{\theta\theta}p_\theta T_\theta 
+ f^{\phi\phi}p_\phi T_\phi
= \frac {p_r}{m} \,, \label{cnopt1}
\end{aligned}
\end{equation}
and
\begin{eqnarray}
\chi_{(4)} &=& \frac{1}{2}\left[ Q^{rr}p_r^2+ Q^{\theta\theta}p^2_\theta + Q^{\phi\phi}p^2_\phi \right] - lw  
\nonumber\\
&=&  \frac{1}{m}  \left[
\frac{p^2_\theta}{mr^3} + \frac{p^2_\phi \sin\theta}{m(b + r\sin\theta)^3} - l
\right]
\label{cnopt2}
\,,
\end{eqnarray}
where we have used $w=m^{-1}$, as can be checked from (\ref{w}) to be the case for the current example.  In equation (\ref{cnopt2}) above we have of course $Q^{rr}=0$ while $Q^{\theta\theta}$ and $Q^{\phi\phi}$ are given by the two non-null entries in (\ref{Qijtorus}).  In passing, we notice that $w=m^{-1}$ fulfills the condition (\ref{assumption}) for a genuine second-class system.
By computing the Poisson brackets between all constraints, the constraint matrix $\Delta_{(rs)} = \{\chi_{(r)}, \chi_{(s)}\}$ can be written as
\begin{equation}
\Delta_{(rs)} =m^{-1}
\begin{bmatrix}
\phantom{-}0 & \phantom{-}0 & 0 & {1} \\
\phantom{-}0 &\phantom{-} 0 & {1} & 0 \\
\phantom{-}0 & -{1} & 0 & \displaystyle\frac{3}{m^3}\Big\{\frac{p^2_\theta}{r^4} + \frac{p^2_\phi {\sin^2\theta}}
{(b + r\sin\theta)^4} \Big\}  \\
-{1}&\phantom{-} 0 & - \displaystyle\frac{3}{m^3}\Big\{\frac{p^2_\theta}{r^4} + \frac{p^2_\phi {\sin^2\theta}}
{(b + r\sin\theta)^4} \Big\}  & 0
\end{bmatrix}
\end{equation}
and matches (\ref{CM}) with entries given by equations (\ref{D}) to (\ref{R}).

Pursuing the Abelianization of the constraints algebra, we introduce four BFFT auxiliary variables  $\eta^{(r)}$, with $r=1,\dots,4$.  Since in this case we have only one initial function $T_\alpha(q^k)$ given by relation (\ref{T=r-a}), i.e., $M=1$, we do not need the extra index $\alpha$.  The corresponding Poisson brackets for the BFFT variables are chosen, according to (\ref{omegars}), as
\begin{equation}
\omega^{rs} \equiv  \{ \eta^{(r)}, \eta^{(s)} \}  =
\begin{bmatrix}
\phantom{-}0 &\phantom{-} 0 & \phantom{-}1&\phantom{-} 0\phantom{-} \\
\phantom{-}0 & \phantom{-}0 &\phantom{-} 0 & \phantom{-}1\phantom{-} \\
-1 & \phantom{-}0 & \phantom{-}0 &\phantom{-} 0 \phantom{-} \\
\phantom{-}0 & -1 &\phantom{-} 0 &\phantom{-} 0\phantom{-}
\end{bmatrix}
\,.
\end{equation}
The functions $T$, $f^{rr}$, $f^{\theta\theta}$, $Q^{rr}$, and $Q^{\theta\theta}$ obtainted from equations (\ref{T=r-a}), (\ref{fij}) and (\ref{Qijtorus}), can be extended to fulfill condition (\ref{c1}) by solving the relations (\ref{r1}) as in (\ref{recsol}).
Specifically, using the recursive relations (\ref{recsol}) and the expansion (\ref{Fbar}), we compute
\begin{equation}
\bar{T}=r-a+\eta^{(2)}
\,,
\end{equation}
\begin{equation}
\bar{f}^{rr}=f^{rr}=m^{-1}\,,\,\,\,\,\bar{Q}^{rr}=Q^{rr}=0\,,
\end{equation}
\begin{equation}
\bar{f}^{\theta\theta}=m^{-1}(r+\eta^{(2)})^{-2}
\,
\end{equation}
and
\begin{equation}
\bar{Q}^{\theta\theta}=\frac{2}{m^2r^3}
\left[
1 - 3\frac{\eta^{(2)}}{r}+6\left(\frac{\eta^{(2)}}{r}\right)^2-10\left(\frac{\eta^{(2)}}{r}\right)^3+\,\,\dots
\right]
=\frac{2}{m^2(r+\eta^{(2)})^3}
\,.
\end{equation}
The same can be said about the partial derivatives of $T$, leading to the extensions
\begin{equation}
 \bar T_r=1\,\,\,\mbox{ and }\,\,\bar T_\theta = \bar T_\phi =0 \,.
\end{equation}
Concerning $f^{\phi\phi}$ and $Q^{\phi\phi}$, the $\theta$-dependence does not allow for a strict solution of (\ref{c1}).  However, if we insist on the recursive expression (\ref{recsol}), we obtain
\begin{equation}
\bar{f}^{\phi\phi}=m^{-1}[b + (r+\eta^{(2)}) \sin\theta]^{-2}
\end{equation}
and
\begin{eqnarray}
\bar{Q}^{\phi\phi}&=&\frac{2\sin\theta}{m^2(b+r\sin\theta)^3}
\left[
1-3\left(\frac{\eta^{(2)}\sin\theta}{b+r\sin\theta}\right)
+6\left(\frac{\eta^{(2)}\,\sin\theta}{b+r\sin\theta}\right)^2-\,\,\dots
\right]
\nonumber\\
&=&\frac{2\,\sin\theta}{m^2\left[b+(r+\eta^{(2)})\sin\theta\right]^3}
\end{eqnarray}
which, from expressions (\ref{chitilde}) and (\ref{chitilde4}), lead to the Abelianized constraints
\begin{equation}\label{cnopt123}
\begin{aligned}
\tilde\chi_{(1)}&=\pi - \eta^{(3)}\,,\\
\tilde\chi_{(2)}&= r-a + \eta^{(2)}\,,\\
\tilde\chi_{(3)}
&=\frac{1}{m}(p_r - \eta^{(4)})\,,
\end{aligned}
\end{equation}
and
\begin{eqnarray}
\tilde\chi_{(4)}& = &\frac{1}{2}\{\bar Q^{\theta\theta}{p^2_\theta} + \bar Q^{\phi\phi}p^2_\phi\}  - \frac{\,\,\,\,l+\eta^{(1)}}{m}\,\,,
\nonumber\\&=&
\frac{1}{m}  \Big\lbrace
\frac{p^2_\theta}{m(r+\eta^{(2)})^3} + \frac{p^2_\phi \sin\theta}{m[b + (r+\eta^{(2)})\sin\theta]^3} - l - \eta^{(1)}
\Big\rbrace\,\,.
\label{cnopt}
\end{eqnarray}
The Poisson bracket between any of the modified constraints above in (\ref{cnopt123}) and (\ref{cnopt}) can be easily checked to identically vanish, that is,
\begin{eqnarray}
\{\tilde \chi_r, \tilde \chi_s\} = 0\,,\,\,\,\,\,r,s=1,\dots,4.
\end{eqnarray}

After obtaining the converted Abelian first-class constraints (\ref{cnopt123}) and (\ref{cnopt}), we next turn our focus to the involutive Hamiltonian.  From the general expression (\ref{Htilde}), using further
$
\bar V = 0
$,
we can write the corresponding modified Hamiltonian as
\begin{eqnarray}\label{modH}
\tilde H& =& \frac{1}{2m}
\Big\lbrace
(p_r-\eta^{(4)})^2+\frac{p_\theta^2}{(r+\eta^{(2)})^2}
+\frac{p_\phi^2}{[b+(r+\eta^{(2)})\sin\theta]^2}
\Big\rbrace
\nonumber\\&&
+(l+\eta^{(1)})(r+\eta^{(2)}-a)
\,.
\end{eqnarray}
It can be immediately checked that (\ref{modH}) is in full involution with the constraints (\ref{cnopt123}).  However, concerning the last constraint (\ref{cnopt}) we have
\begin{equation}
\lbrace \tilde{\chi}_{(4)}, \tilde{H}  \rbrace
=\frac{3bp_\theta p_\phi^2\cos\theta}{m^3(r+\eta^{(2)})^2
[b+(r+\eta^{(2)})\sin\theta]^4}\,.
\end{equation}
This can be understood by the fact already mentioned that $\bar{f}^{\phi\phi}$ and $\bar{Q}^{\phi\phi}$ are not strict solutions to (\ref{c1}).  In order to obtain involution with $\tilde{\chi}_{(4)}$, we may improve the modified Hamiltonian with $\eta^{(3)}$-higher-order terms.  Namely, redefinining
\begin{equation}
\tilde{\tilde{H}}={\tilde{H}}
+m\eta^{(3)} \lbrace \tilde{\chi}_{(4)}, \tilde{H}  \rbrace
+\frac{m^2{\eta^{(3)}}^2}{2} \lbrace \tilde{\chi}_{(4)},
\lbrace \tilde{\chi}_{(4)}, \tilde{H}  \rbrace
\rbrace
\,+\,\,\dots
\end{equation}
we obtain an extended Hamiltonian $\tilde{\tilde{H}}$
which can be checked to be have a null Poisson bracket with all converted constraints, forming a complete Abelianized involutive algebra.
We have thus achieved our goal of converting the second-class system (\ref{lagft}) to an Abelian first-class involutive one.  In the next section, we show an example of application of this Abelianization scheme to a field theory model.

\section{Field Theory Example}
The prototypical system (\ref{L}) can describe interesting field theory models.  We have already mentioned, for instance, the nonlinear sigma model which has been recently studied along similar lines in reference \cite{Thibes:2020yux}.  Here we shall discuss a new application to field theory, namely the Lorentz symmetry breaking bumblebee model \cite{Nambu:1968qk, Kostelecky:2003fs, Bluhm:2006im}.  Consider a scalar and a vector fields $\phi$ and $B_\mu$ interacting via the action functional
\begin{equation}\label{S}
S[\phi,B^\mu]=-\frac{1}{2}\int d^Dx\,
\left[
\partial_\mu B_\nu \partial^\mu B^\nu
+\phi\left(B_\mu B^\mu - a^2 \right)
\right]
\,,
\end{equation}
where $a$ denotes a given non-null real parameter.  For definiteness, we work in a $D$-dimensional Minkowski space with flat metric $\eta^{\mu\nu}$, signature $(+,-,\dots,-)$, but the whole analysis could be carried out with any other signature, including Euclidian space as well, just watching for a few signs change. The Lorentz indexes, denoted by Greek letters, run from $0$ to $D-1$.  The scalar field $\phi$ acts as a Lagrange multiplier imposing the condition
\begin{equation}
B_\mu B^\mu = a^2
\,,
\end{equation}
which gives rise to a spontaneous Lorentz symmetry breaking.  Due to the non-null expectation value for $B_\mu B^\mu$, the bumblebee vector field chooses a preferred specific direction along the Minkowski space characterizing thus a spontaneous breaking of the Lorentz symmetry.  The Lagrangian density associated to (\ref{S}) can be cast into the form (\ref{L}) as
\begin{equation}
{\cal L}=-\frac{1}{2}\eta_{\mu\nu}\dot{B}^\mu\dot{B}^\nu
-{\cal V}(B^\mu)
-\frac{1}{2}\phi\left(B_\mu B^\mu - a^2 \right)
\,,
\end{equation}
with $M=1$ and potential density function
\begin{equation}\label{V(B)}
{\cal V}(B^\mu)\equiv
-\frac{1}{2}\nabla B_\mu\cdot\nabla B^\mu
\,.
\end{equation}
The nabla operator $\nabla$ in (\ref{V(B)}) stands for the gradient in the $D-1$ space coordinates. In accordance with our previous notations, we define
\begin{equation}
T\equiv \frac{1}{2}\left(B_\mu B^\mu - a^2 \right)
\end{equation}
with corresponding derivatives with respect to $B_\mu$ written as\footnote{Boldface denotes the $(D-1)$-spatial part of a contravariant $D$-vector, so that $x^\mu\equiv(x^0,\mathbf{x})$ etc.} 
\begin{equation}
\label{Tmu}
T_\mu(x,y)|_{y^0=x^0}=B_\mu(x)\delta^{(D-1)}(\mathbf{x}-\mathbf{y})
\,.
\end{equation}

Introducing the canonical momenta $\pi_\phi$ and $\Pi_\mu$, respectively conjugated to the fields $\phi$ and $B^\mu$, we may perform a Legendre transformation and obtain the canonical Hamiltonian
\begin{equation}\label{75}
H=\int\,d^{D-1}x
\left[
\frac{1}{2}{(\Pi^i)}^2-\frac{1}{2}{(\Pi^0)}^2
+{\cal V}(B^\mu)
+\frac{1}{2}\phi\left(B_\mu B^\mu - a^2 \right)
\right]
\,,
\end{equation}
which corresponds to equation (\ref{H}) with opposite Minkowski metric, namely
\begin{equation}\label{feta}
f^{ij}(q^k)\longrightarrow-\eta^{\mu\nu}\delta^{(D-1)}(\mathbf{x}-\mathbf{y})
\,.
\end{equation}
For the present case then, using (\ref{Tmu}), equations (\ref{Qij}) to (\ref{w}) lead to relations
\begin{equation}\label{Qmunu}
Q^{\mu\nu}(x,y,z)|_{z^0=y^0=x^0}=2\eta^{\mu\nu}\delta^{(D-1)}(\mathbf{x}-\mathbf{y})\delta^{(D-1)}(\mathbf{y}-\mathbf{z})
\,,
\end{equation}
\begin{equation}
v = - B_\mu \nabla^2 B^\mu
\,,
\end{equation}
and
\begin{equation}
w(x,y)|_{y^0=x^0}=-B_\mu(x) B^\mu(x) \, \delta^{(D-1)}{(\mathbf{x}-\mathbf{y})}
\,,
\end{equation}
from which we obtain a chain of four second-class constraints
\begin{equation}\label{Bconsts}
\begin{aligned}
{\chi}_{(1)}&=\pi_\phi
\,,\\
\chi_{(2)}&=\frac{1}{2}\left(B_\mu B^\mu - a^2 \right)
\,,\\
{\chi}_{(3)}&=-\Pi_\mu B^\mu
\,,\\
{\chi}_{(4)}&= \Pi_\mu \Pi^\mu + \phi B_\mu B^\mu + B_\mu
\nabla^2 B^\mu
\,,
\end{aligned}
\end{equation}
corresponding to equations (\ref{Xi1}) and (\ref{Xi2}) to (\ref{Xi4}).  Alternatively, the above constraint relations (\ref{Bconsts}) can be checked to be generated by direct application of the Dirac-Bergmann procedure.

Hence, we see that action (\ref{S}) describes a second-class constrained dynamical system with no gauge-freedom.  Nonetheless, it is possible to bypass the need for Dirac brackets and perform its functional quantization, for instance along the BFV lines, insofar as we convert its constraints from second- to first-class, as we have discussed in general terms in Section {\bf 3}.
For that matter, pursuing the Abelianization of the constraints (\ref{Bconsts}), we introduce four BFFT fields $\eta^{(r)}$, with $r=1,\dots,4$, satisfying the equal-time non-null Poisson bracket relations
\begin{equation}
\{\eta^{(1)}({x}),\eta^{(3)}({y})\}|_{y^0=x^0}
=
\{\eta^{(2)}({x}),\eta^{(4)}({y})\}|_{y^0=x^0}
=
\delta^{(D-1)}(\mathbf{x} - \mathbf{y})
\,.
\end{equation}
In this case, since (\ref{feta}) and (\ref{Qmunu}) do not depend on $B^\mu$, we have the immediate trivial solution
\begin{equation}
\bar Q^{\mu\nu}(x,y,z)|_{z^0=y^0=x^0} = Q^{\mu\nu}(x,y,z)|_{z^0=y^0=x^0}
\end{equation}
for equation (\ref{c1}).  However, concerning the second equation in (\ref{fbar}), it is not possible to find an extension of (\ref{Tmu}) satisfying the condition (\ref{c1}).   Be that as it may, it is still possible to obtain a constraints Abelian conversion in closed form, namely,
\begin{equation}\label{conv}
\begin{aligned}
\tilde{\chi}_{(1)}=&\,\pi_\phi-\eta^{(3)}
\,,\\
\tilde\chi_{(2)}=&\,\frac{1}{2}\left(B_\mu B^\mu - a^2 \right)+\eta^{(2)}
\,,\\
\tilde{\chi}_{(3)}=&\,-\left(\Pi_\mu B^\mu-B_\mu B^\mu\eta^{(4)}\right)
\left(1+2(B_\lambda B^\lambda)^{-1}\eta^{(2)}\right)^{1/2}
\,,
\end{aligned}
\end{equation}
and
\begin{eqnarray}\label{conv4}
\tilde{\chi}_{(4)}&=& \frac{{[B_\mu(\Pi^\mu-B^\mu\eta^{(4)})]}^2}{B_\lambda B^\lambda}
\,+\,
\frac{B_\nu B^\nu \Pi_\mu \Pi^\mu - B_\mu\Pi^\mu B_\nu \Pi^\nu}
{B_\rho B^\rho \left[1+2(B_\lambda B^\lambda)^{-1}\eta^{(2)}\right]}
\nonumber\\
&&+\,\left(1+2(B_\lambda B^\lambda)^{-1}\eta^{(2)}\right)^{1/2}
B_\mu \, \nabla^2
\left[
\left(1+2(B_\lambda B^\lambda)^{-1}\eta^{(2)}\right)^{1/2}B^\mu
\right]
\nonumber\\&&
+\,(\phi+\eta^{(1)})\left(1+2(B_\lambda B^\lambda)^{-1}\eta^{(2)}\right)B_\mu B^\mu
\,.
\end{eqnarray}
Note that the four converted constraints above clearly reproduce the original second-class ones (\ref{Bconsts}) in the limit $\eta^{(r)}\rightarrow0$.  Additionally,
by using the fact that
\begin{equation}\label{84}
\bigg\lbrace 
\left(1+\frac{\,\,2\eta^{(2)}}{(B_\lambda B^\lambda)}\right)^{1/2}\!\!\!\!\!B^\mu
\,,\,
\Pi_\nu-B_\nu\eta^{(4)}
\bigg\rbrace
=
\left(1+\frac{\,\,2\eta^{(2)}}{(B_\lambda B^\lambda)}\right)^{1/2}
\left[\delta^\mu_\nu-\frac{B^\mu B_\nu}{B_\lambda B^\lambda}\right]
\,,
\end{equation}
the constraints  (\ref{conv}) and (\ref{conv4})  can be explicitly checked to satisfy the Abelian condition
\begin{eqnarray}
\{\tilde \chi_r, \tilde \chi_s\} = 0\,,\,\,\,\,\,r,s=1,\dots,4.
\end{eqnarray}

As for the involutive modified Hamiltonian $\tilde H = \tilde H(B_\mu,\phi,\Pi^\mu,\pi_\phi,\eta^{(r)})$, we write
\begin{eqnarray}\label{miH}
\tilde H &=&
\int\,d^{D-1}x
\left[
\frac{{[B_\mu(\Pi^\mu-B^\mu\eta^{(4)})]}^2}{2B_\lambda B^\lambda}
\,+\,
\frac{B_\nu B^\nu \Pi_\mu \Pi^\mu - B_\mu\Pi^\mu B_\nu \Pi^\nu}
{2B_\rho B^\rho \left[1+2(B_\lambda B^\lambda)^{-1}\eta^{(2)}\right]}
\right.\nonumber\\&&\left.
+\frac{1}{2}(\phi+\eta^{(1)})\left(B_\mu B^\mu - a^2 \right)+(\phi+\eta^{(1)})\eta^{(2)}
+\bar{\cal V} (B^\mu,\eta^{(2)})
\right]
\,,
\end{eqnarray}
with modified potential function
\begin{equation}
\bar{\cal V} (B^\mu,\eta^{(2)})\equiv-
\nabla
\left[
\left(1+2(B_\lambda B^\lambda)^{-1}\eta^{(2)}\right)^{1/2}B_\mu
\right]
\cdot
\nabla
\left[
\left(1+2(B_\lambda B^\lambda)^{-1}\eta^{(2)}\right)^{1/2}B^\mu
\right]
\,.
\end{equation}
We can see by inspection that $\tilde H$ above satisfies the boundary condition
\begin{equation}
 \tilde H(B_\mu,\phi,\Pi^\mu,\pi_\phi,0) = H(B_\mu,\phi,\Pi^\mu,\pi_\phi)
\end{equation}
with the original $H$ given by (\ref{75}) and, using again (\ref{84}), that
\begin{equation}
\{  {\tilde\chi}_{(r)}, \tilde{H} \} = 0\,,\,\,\,r=1,\dots,4
\,.
\end{equation}

In this way, we have shown how the prototypical system (\ref{L}) can be used in a field theory context.  The summations in repeated indexes have been generalized in a DeWitt notation fashion to properly include spatial integrations when necessary.
Although the conditions (\ref{c1}) were not all strictly satisfied, we were able to 
 successfully obtain an Abelian involutive algebra for the converted constraints and Hamiltonian in closed form for the spontaneous Lorentz symmetry breaking bumblebee model (\ref{S}).

\section{Conclusion}
The interplay between first- and second-class constraints has been analyzed in many different models in the literature, both at classical and quantum levels, with respect to various aspects regarding their constraints structure.
The knowledge of that structure is essential when it comes to understanding the gauge algebra, BRST symmetries, ghost field properties, anomalies and renormalizability.  It also plays a fundamental role in important alternative approaches such as supersymmetry, superfield formulations and generalizations of BRST transformations.
As we have pointed out in the Introduction, it is long known that the quantization of second-class dynamical systems can be performed via a constraints Abelianization program, as is clearly attested by the large amount of published works in that direction dealing with many specific physical models.  However, it is not difficult to see that many of these models enjoy similarities among themselves, sometimes hidden by different contexts, notations and conventions.  Even mechanical models can resemble particular quantum field theory aspects, possessing very much the same structure when it comes to their constraint relations.  We have shown that it is possible to capture the main key aspects of a broad class of such models into a common prototypical system for which we have derived very handy and useful general relations regarding its constraints structure.  For this prototypical system, furthermore, we have derived clear sufficient conditions to achieve the constraints abelianization and involutive Hamiltonian in a simple closed form, obtaining in general terms a corresponding first-class involutive system for that case.
It is important to stress that the prototypical system discussed here is by no means restricted to linear constraints.

We have seen that the prototypical second-class system (\ref{L}) is able to comprehensively capture the essential properties of seemingly different physical situations.  Naturally this brings the possibility of highly unifying the study of such specific models.   We mention the non-linear sigma model \cite{Thibes:2020yux}, the quantum rigid rotor \cite{Rai:2010aa}, the generalized quantum rigid rotor \cite{Thibes:2020yux, deOliveira:2019eva} and the conic constrained particle \cite{Barbosa:2018dmb, Barbosa:2018bng} as instances of second-class systems which could be treated along the same lines as the particle on a torus and bumblebee Lorentz symmetry breaking models discussed here.   The bumblebee model enjoys a rich physical structure, being able to promote a spontaneous breaking of the Lorentz symmetry, it can be used as a building block for a more ambitious description of our universe \cite{Kostelecky:2003fs} -- nonetheless its essential constraint structure as a dynamical system can be described by the simple prototypical system (\ref{L}).  Initially seen as a second-class system, we have explicitly obtained the complete Abelianization for the bumblebee model (\ref{S}) in a closed form.
The extension and application of our results to more complex and intricate situations, aiming at an even larger unification, is currently under analysis.

\end{document}